\documentclass[12pt,a4paper]{article}
\usepackage[utf8]{inputenc}
\usepackage{amsmath}
\usepackage{amssymb}
\usepackage{graphicx}
\usepackage{subfigure}
\usepackage[affil-it]{authblk}
\usepackage{etoolbox}
\usepackage{lmodern}

\makeatletter
\patchcmd{\@maketitle}{\LARGE \@title}{\fontsize{16}{19.2}\selectfont\@title}{}{}
\makeatother

\author[1,2]{Rafael A. Bizao}
\author[1]{Tiago Botari}
\author[1,3]{Eric Perim}
\author[2,4,5]{Nicola M. Pugno}
\author[1,*]{Douglas S. Galvao}
\affil[1]{Instituto de F\'isica Gleb Wataghin, Universidade Estadual de Campinas, 13083-970, Campinas, SP, Brazil.}
\affil[2]{Department of Civil, Environmental and Mechanical Engineering, Laboratory of Bio-Inspired and Graphene Nanomechanics, University of Trento, via Mesiano, 77, 38123 Trento, Italy.}
\affil[3]{Department of Mechanical Engineering and Materials Science, Duke University, Durham, NC 27708, USA.}
\affil[4]{Italian Space Agency, Via del Politecnico snc, 00133 Rome, Italy.}
\affil[5]{School of Engineering and Materials Science, Queen Mary University of London, Mile End Road, London E1 4NS, United Kingdom.}
\affil[*]{galvao@ifi.unicamp.br}
\date{}

\title{Mechanical Properties and Fracture Patterns of Graphene (Graphitic) Nanowiggles}

\begin{document}
\maketitle

\begin{abstract}
Graphene nanowiggles (GNW) are graphene-based nanostructures obtained by making alternated regular cuts in pristine graphene nanoribbons. GNW were recently synthesized and it was demonstrated that they exhibit tunable electronic and magnetic properties by just varying their shape. Here, we have investigated the mechanical properties and fracture patterns of a large number of GNW of different shapes and sizes using fully atomistic reactive molecular dynamics simulations. Our results show that the GNW mechanical properties are strongly dependent on its shape and size and, as a general trend narrow sheets have larger ultimate strength and Young's modulus than wide ones. The estimated Young's modulus values were found to be in a range of $\approx 100-1000$ GPa and the ultimate strength in a range of $\approx 20 - 110$ GPa, depending on GNW shape. Also, super-ductile behaviour under strain was observed for some structures.
\end{abstract}

%\flushbottom

\section*{Introduction}

Graphene is a carbon allotrope obtained by arranging carbon atoms on two-dimensional (2D) honeycomb lattice. The advent of graphene \cite{Geim2007,Novoselov2004} created a revolution in materials science, due to its unique and exceptional electronic and mechanical properties. Because of these properties, graphene has great potential for applications in different fields, such as energy storage \cite{Mukherjee2012,Stoller2008}, solar cells \cite{doi:10.1021/nn304410w} and nanoelectronics \cite{Schwierz2010}. However, in its pristine form, graphene is a zero band gap semiconductor, which poses limitations to its use in applications such as digital transistors. There are several ways to open graphene band gap, including chemical functionalization \cite{PhysRevB.78.085413}, application of mechanical stress \cite{PhysRevB.85.125403} or by topological structural changes, notably the synthesis of narrow strips called graphene nanoribbons (GNR) \cite{Terrones2010351}. GNR can be defined as finite graphene segments with large aspect ratio. Their electronic properties have been extensively studied \cite{Baringhaus2014,Han2007} and shown to be directly related to electron confinement arising from constraints due to finite boundaries. In this way, as GNR becomes narrower the band gap increases, lowering the conductance \cite{Han2007}.

Recently, with the report of a precise bottom-up fabrication technique \cite{Cai2010}, it was possible to synthesize GNR in an easier and more controlled way when compared to other methods, such as chemical vapor deposition \cite{Campos-Delgado2008} and unzipping of carbon nanotubes \cite{doi:10.1021/ja203860a}. This new method allows the synthesis of not just rectangular structures but also different GNR shapes called graphene or graphitic nanowiggles (GNW) \cite{Cai2010,giraoprl}. It uses different types of monomers as molecular precursors in a surface-assisted coupling method \cite{Grill2007,Gourdon2008}. The resulting GNW shape depends on the structure of the precursor monomer, which is easy to control. This enables the experimental synthesis and systematic study of GNW of different shapes.

\begin{figure}
\centerline{\includegraphics[width=0.8\linewidth]{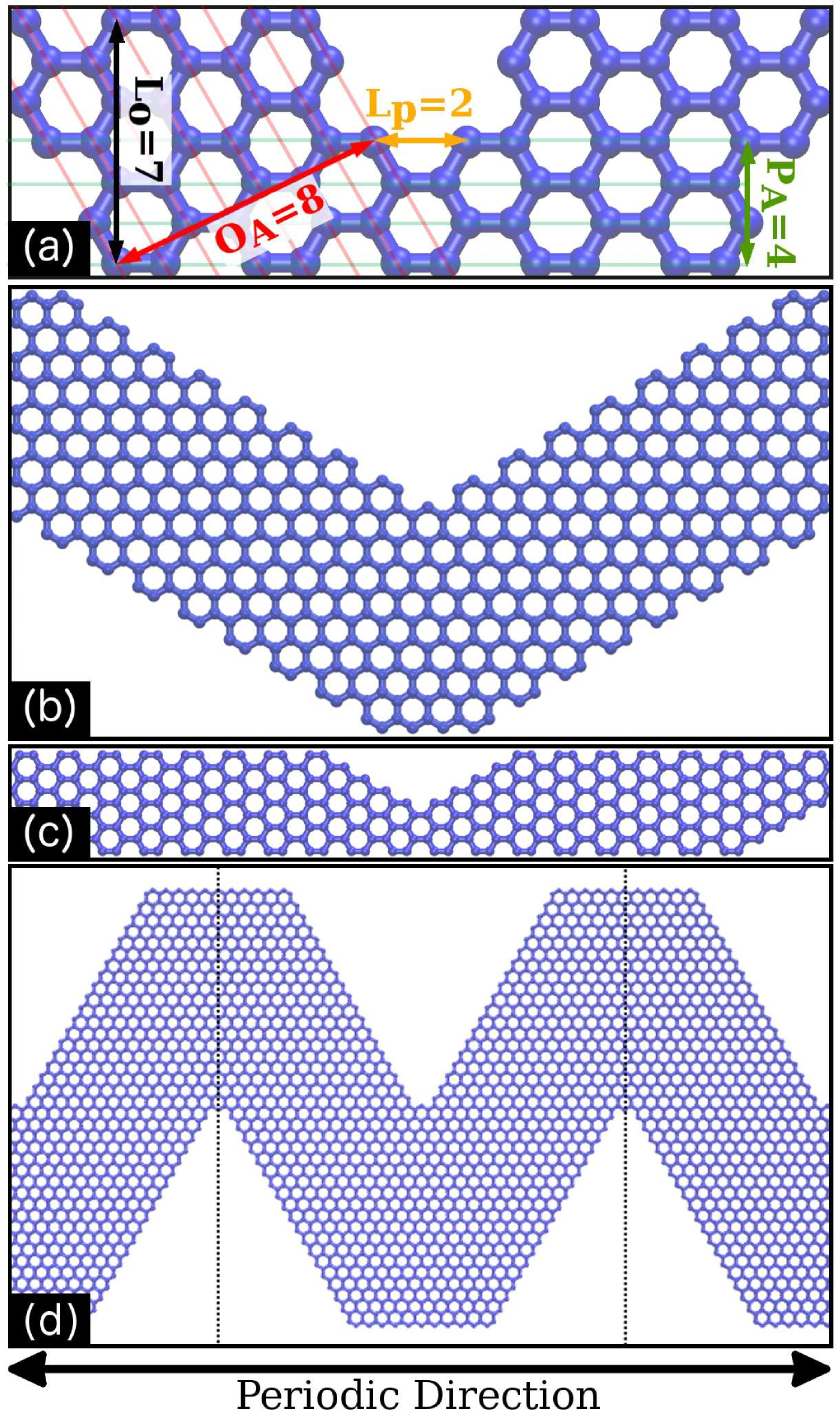}}
\caption{Examples of GNW structures. (a) the main four structural parameters ($P_\alpha$, $O_\beta$,  $L_p$ and $L_o$) used to uniquely define the GNW structures: ($P_\alpha$, $O_\beta$,  $L_p$, $L_o$) = (a) $(4_A,8_A,2,7)$; (b) $(9_Z,15_A,1,17)$; (c) $(5_A,10_Z,2,9)$ and; (d) $(20_Z,15_Z,1,39)$. In (d) the dotted lines indicate the structural unit cell. For more details about GNW definition and structural characterizations, see reference \cite{giraoprb}.}
\label{estruturas}
\end{figure}

Basically, GNW consist of non-aligned periodic repetitions of GNR, in a chevron-type graphene nanoribbon structure, as shown in figure \ref{estruturas}. The full description of the shape of a GNW depends on four structural parameters \cite{bizao2014mechanical}: the width of the structure, $L_o$ (measured perpendicular to its length) and the length of the oblique, $O_\beta$, and outer/inner parallel, $P_\alpha$/$L_p$, segments. Oblique and parallel directions are defined with respect to the length direction of the structure (\textit{i.e.}, its longest direction). These parameters are illustrated in Fig. \ref{estruturas}(a). The $\alpha$/$\beta$ sub-index denotes the morphology of the parallel/oblique segment, either armchair ($A$) or zig-zag ($Z$). Under this representation, four different GNW families may be defined: $(\alpha,\beta)=(A,A)$ (Fig. \ref{estruturas}(a)), $(Z,A)$ (Fig. \ref{estruturas}(b)), $(A,Z)$ (Fig. \ref{estruturas}(c)) and $(Z,Z)$ (Fig. \ref{estruturas}(d)).

It was recently shown that GNW present electronic and magnetic properties that can be tuned just by changing their shape \cite{giraoprl}. This enables the tailoring of these structures for specific applications. Although their electronic and spintronic properties have been studied in detail, the study of their mechanical properties and fracture patterns under strain is still missing. 

The detailed knowledge of the mechanical properties of these materials is very important for the fabrication of nanodevices and for the exploitation of their graphene-like strength \cite{Ovid'ko2013,Faccio2009}. The mechanical properties of graphene have been intensively investigated by different methodologies, both in experiments and in theory. Based on atomic force microscope (AFM) nanoindentation experiments, it was found that the Young's modulus of free standing graphene sheets could reach values as high as $1.0$ TPa and the critical stress (also known as tensile strength and as ultimate strength) could reach $130$ GPa \cite{Lee2008}, which makes graphene the strongest material ever measured. Computer simulations using {\it ab-initio} calculations through density functional theory (DFT) are consistent with these results, obtaining $1.05$ TPa for the Young's Modulus and $130$ GPa for the critical stress \cite{Liu2007}. More recently, it was shown that the GNR mechanical properties depend on their width \cite{Faccio2009,Topsakal2010,Bu2009}, and that they can be harder than graphene and graphene nanotubes due to edge reconstruction effects \cite{Faccio2009}.

However, these remarkable mechanical properties are very sensitive to defects. Even in small amounts they significantly decrease the Young's modulus and the ultimate strength values \cite{Banhart2011}. In the case of grain boundaries, the mechanical properties remain almost unchanged from the pristine graphene sheet \cite{Lee2013}, in agreement with previous theoretical predictions \cite{Grantab946}. However, when vacancies and Stone-Wales defects are considered, the Young's modulus and ultimate strength decrease with the increase in the density of defects, reaching a saturation point in the high-ratio regime \cite{Xu2013}. Additionally, the insertion of defects in graphene leads to changes in their fracture patterns, transitioning from brittle to super-ductile behavior \cite{Xu2013}.

GNW represent an attempt to tune the mechanical properties of graphene-like materials using structures that do not exhibit vacancies and/or Stone-Walles-like defects, although they can be considered themselves as a kind of graphene with topological defects.

In this work, we present a thorough and systematic investigation of the GNW mechanical properties and fracture dynamics. A large number of distinct GNW shapes and sizes were investigated, comprising all four different families. Using reactive molecular dynamics simulations we calculate the Young's modulus, the ultimate strength, the von Mises stress distribution and the fracture patterns for over $1000$ unique structures. Two different temperatures, $10K$ and $300K$, were used in order to investigate their temperature dependence. Our results show that the GNW mechanical response can be tuned to a large range of values, while being very sensitive to $P_\alpha$, $O_\beta$,  $L_p$ and $L_o$ values. This tunable mechanical behaviour associated with tunable electronic and magnetic properties \cite{giraoprl}, makes GNW very attractive structures to be exploited as advanced functional materials.

\section{Methodology}
\label{methodology}

The present study was carried out through molecular dynamics (MD) simulations using the reactive force field ReaxFF \cite{reax}, as implemented in the LAMMPS package \cite{lammps}. Simulations were performed using an accurate timestep of $0.05$ fs at two different temperatures, $10$ K and $300$ K. The temperature values were controlled by a chain of Nos\'e-Hoover thermostats. The results discussed below are for $10$ K unless otherwise stated. Results for $300$ K will be discussed when relevant.

ReaxFF is a general distance-dependent bond order potential in which the van der Waals and Coulomb interactions are explicitly considered \cite{Rappe}. It can reliably describe the formation and dissociation of chemical bonds among atoms, thus allowing the study of chemical reactions. Its use is attractive in cases where the use of \textit{ab initio} methods becomes computationally prohibitive, \textit{i.e.}, for large systems and for long simulation times. The force field parametrization was developed using very accurate DFT calculations and experimental data when available \cite{Rappe}.
ReaxFF has been successfully used in investigations of mechanical properties of silicene membranes \cite{Botari2014}, graphene-like carbon nitride sheets \cite{de2016mechanical}, graphene healing mechanisms \cite{Botari2016302}, combustion \cite{Weismiller2010,Qian2011} and oxidation \cite{Chenoweth2008} of carbon-based systems, etc. In our simulations we adopted Mueller's parametrization \cite{Mueller2010}. This parametrization has been shown to produce good results in the study of mechanical properties of carbon-based nanostructures. 

GNW structures were built with an average of $2000$ atoms. The methods for obtaining GNW unit cells are described in reference \cite{giraoprb}. Herein, $P_\alpha$ and $O_\beta$ are measured as the number of lines of carbon atoms parallel to the respective direction and into the corresponding region. These lines are shown in \ref{estruturas}(a) for both directions. This is equivalent to measuring them as $2n+1$, where $n$ is the number of rings along the perpendicular direction, and $n$ can assume semi-integer values. $L_o$ is measured similarly to $P_\alpha$, however, it accounts for all lines of carbon atoms that are parallel to the GNW length, including the ones within the oblique region. The same formula using the number of rings applies to $L_o$. $L_p$ is measured as the number of carbon atoms in the innermost parallel segment that have only 2 nearest-neighbors. For all structures the $L_p$ parameter was taken as the smallest possible,  \textit{i.e.}, $L_p=2$ if $\alpha=A$ or $L_p=1$, if $\alpha=Z$. The $L_o$ parameter was chosen such that $L_o=2\times P_\alpha-1$. In this way, the structures can be defined just by the $P_\alpha$ and $O_\beta$ parameters. We excluded the forbidden (in terms of carbon valence) geometries of the combination of $P_\alpha$, $O_\beta$, $L_p$ and $L_o$. 

In order to perform stress/strain calculations, we first carefully thermalized the structures using a NPT ensemble, fixing the external pressure to zero along the periodic directions (see Fig. \ref{estruturas}). The thermalization procedure is performed in order to eliminate any residual stress from thermal effects. After this, we used a NVT ensemble and continuously (until mechanical failure) stretched the structure by applying strain along the periodic directions. Stress values were computed at each time step. We adopted a strain rate of $1\times 10^{-5}$ fs$^{-1}$, which was found to be adequate after several tests.

In order to obtain the stress values, we calculated the virial stress tensor, given by 
\begin{equation} 
\sigma_{ij}=\frac{\sum_{k}^{N}m_{k}v_{k_i}v_{k_j}}{V}+\frac{\sum_{k}^{N}r_{k_i}\cdot f_{k_j}}{V},
\end{equation}
in which $V= l\times A$ is the structure volume, $N$ the number of atoms, $v$ the velocity, $r$ the atom position and $f$ the force per atom. The GNW volume was calculated during the stretching process and the total area at zero strain ($A_0$). $A_0$ was calculated by multiplying the total length and the total width of the GNW and then subtracting the area of the trapezoidal regions that are empty. The total area was assumed to grow linearly with the strain, i. e., $A=(1+\epsilon)A_0$, where $\epsilon$ is the strain. We adopted the thickness of a graphene sheet as being $l = 3.4$ \AA.

The stress-strain curves were obtained by plotting the uniaxial component of the stress tensor ($\sigma_{ii}$) along the periodic direction ($i$) and the strain ($\epsilon_i$), which is defined as a dimensionless quantity dividing the actual deformation by the initial size of the structure along that direction, i. e.:
\begin{equation}
\epsilon_i =\frac{\Delta L_i}{L_i^o}.
\end{equation}
where  $\Delta L_i= L_i-L_i^o$ is the variation along the $i$ direction, $L_i$ is the actual dimension and $L_i^o$ is the initial length of the structure. The Young's modulus values can be obtained as the ratio between the uniaxial stress and the strain applied along the periodic direction at the linear regime
\begin{equation}
Y = \frac{\sigma_{ii}}{\epsilon_i}.
\end{equation}
where $\sigma_{ii}$ is the $ii$ component of the virial stress tensor.

We also calculated the von Mises stress values for each atom in order to obtain information regarding the stress distribution on the strained structure. The von Mises stress provides helpful information on the fracture process, since it is possible to easily visualize its distribution throughout the whole structure \cite{DosSantos2012,de2016mechanical,Botari2014}. The von Mises per atom (i) stress is defined as 
\begin{equation}
\resizebox{.8\hsize}{!}{$\sigma_{vm}^i=\sqrt{\frac{\left(\sigma_{11}^i-\sigma_{22}^i\right)^2+\left(\sigma_{22}^i-\sigma_{33}^i\right)^2+\left(\sigma_{11}^i-\sigma_{33}^i\right)^2
+6\left({\sigma_{12}^i}^{2}+{\sigma_{23}^i}^{2}+{\sigma_{31}^i}^{2}\right)}{2}}$},
\end{equation}
in which the $\sigma_{ii}$ ($i=1,2,3$) and $\sigma_{ij}$ ($i\neq j=1,2,3$) components are the normal and shear stresses, respectively.

\section{Results and Discussions}
\label{results}

\begin{figure}
\centering
\mbox{\subfigure[]{\includegraphics[width=0.73\linewidth]{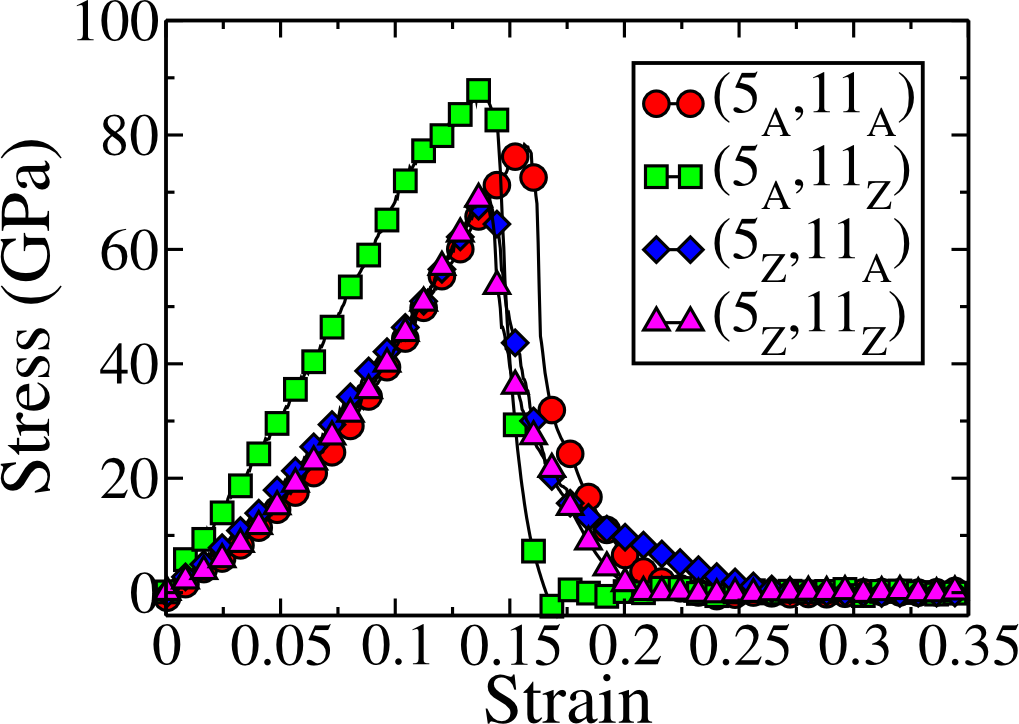}}\quad}\\
\mbox{\subfigure[]{\includegraphics[width=0.73\linewidth]{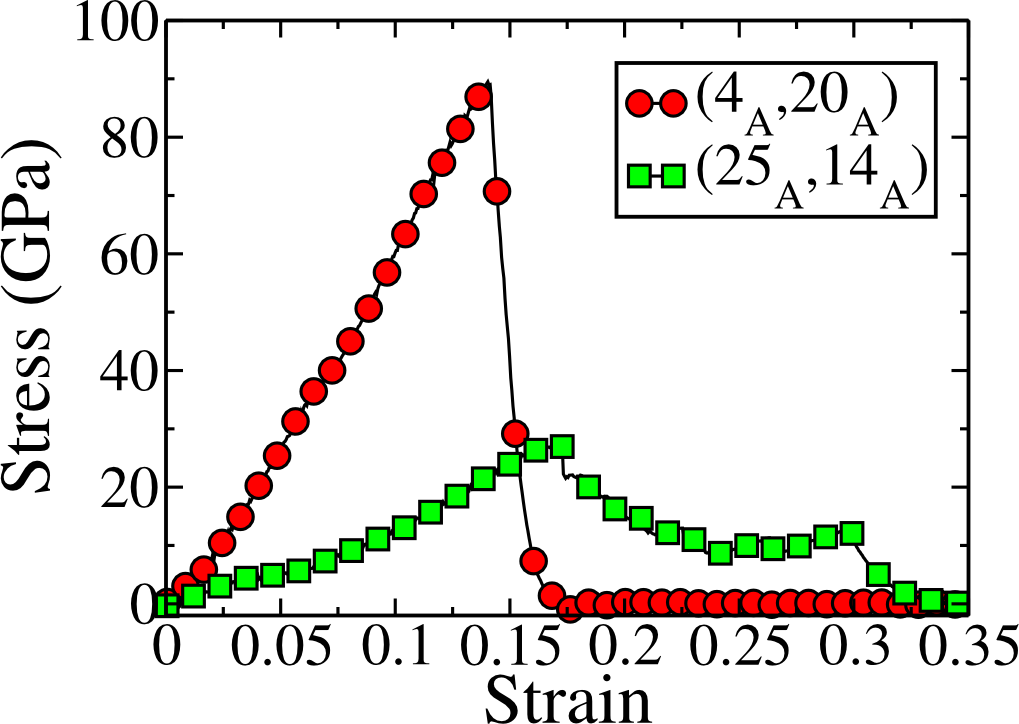}}\quad}
\caption{\textit{Representative stress/strain curves. (a) of all GNW families keeping the $P_\alpha$ and $O_\beta$ indexes constant. The obtained results for Young's modulus ($Y$) and ultimate strength ($\sigma_c$) were $Y=648(2)$ GPa and $\sigma_c\approx 78$ GPa for $(5_A,11_A)$, $Y=763(3)$ GPa and $\sigma_c\approx 89$ GPa for $(5_A,11_Z)$, $Y=659(5)$ GPa and $\sigma_c\approx 69$ GPa for $(5_Z,11_A)$ and $Y=664(4)$ GPa and $\sigma_c\approx 70$ GPa for $(5_Z,11_Z)$; (b) for the AA family showing super-ductility for large $P_\alpha$ values. As can be seen, for a $(25_A$,$14_A)$ GNW fracture behavior is ductile and complete rupture is observed at approximately $30\%$ strain, while for $(4_A$,$20_A)$ GNW fracture behavior is brittle and complete rupture is observed at approximately $15\%$ strain.}}
\label{stressall}
\end{figure}

Firstly, we analyzed the relaxed GNW structures at finite temperatures. These structures are obtained through a thermalization process, as discussed in the methodology section. After these processes, GNW exhibit structural corrugations, as illustrated in Fig. 1 of the supplementary material. The level of corrugation depends on the GNW family ($AA$, $AZ$, $ZA$ or $ZZ$) and on the values of $P_\alpha$ and $O_\beta$. The $AA$ family presents the smallest levels of corrugation, assuming considerable values only for structures with high $P_\alpha$ values. For the $AZ$ family, the corrugation levels can be considerable for large values of $P_\alpha$ and small values of $O_\beta$, but is very small for small values of $P_\alpha$. Considering the $ZA$ family, the corrugation level increases for large values of $O_\beta$. Finally, for the $ZZ$ family, the corrugation level becomes small only for structures with very small values of $O_\beta$ and $P_\alpha$, assuming significant levels otherwise. These trends can be better visualized in Fig. 2 of the supplementary material, where we present the average quadratic out-of-plane position ($<z^2>$) for the different structures.  As expected, the temperature of the thermalization process also influences the GNW corrugation levels.

%Before the calculations of the mechanical properties, we first analysed the final structures obtained after the thermalization process at NPT ensemble, see methodology section for details. We found out of plane undulations in the final structures (see material supplementary figure 2). The undulation level mainly depend on the family, $P_\alpha$ and $O_\beta$ indexes that generate the structure.
%In the case of $AA$ family, the detected undulations are small and almost insignificant, taking non-zero values only for structures with high $P_\alpha$ values. While for the $AZ$ family, the undulations are considerable and increase for high values of $P_\alpha$ and small values of $O_\beta$. Considering the $ZA$ family, the undulation increases for high values of $O_\beta$ and small values of $P_\alpha$. Finally, for the $ZZ$ family, the undulation becomes very small for structures with small values of $O_\beta$. Another factor that can influence these undulations is the chosen thermalization temperature due to thermal fluctuation changes. The quadratic out of plane position average ($<z^2>$) for different structures after the thermalization process that supports the previous discussion can be found in the supplementary material figure 2. 

\begin{figure}
\centerline{\includegraphics[width=0.7\linewidth]{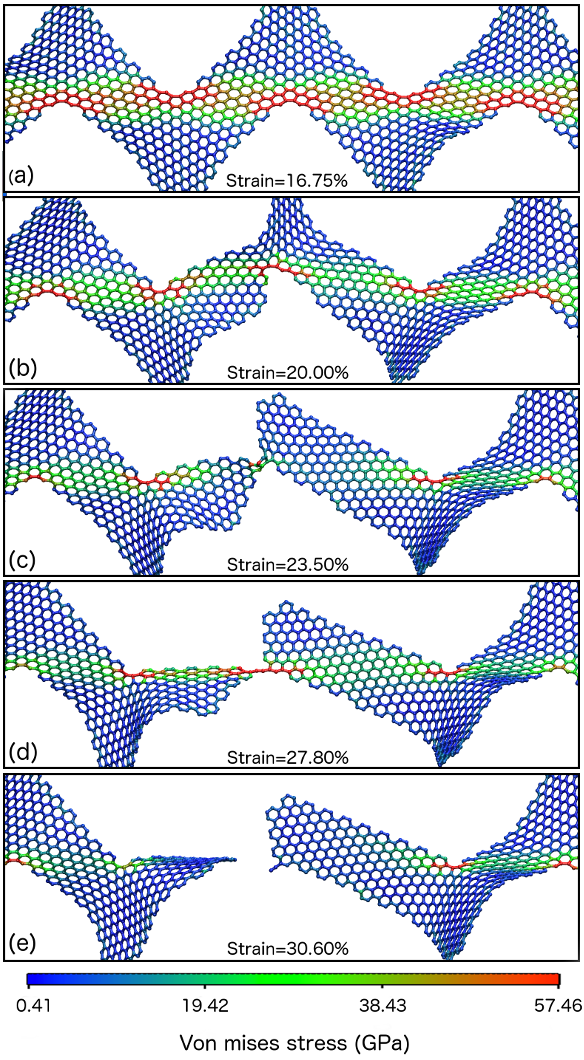}}
\caption{(a-d) Fracture dynamics of the $(25_A,14_A)$ GNW showing the von Mises stress distribution. The stress concentrates on a central line that propagates along the structure. As the strain increases, stress accumulates at the corner region of the structure and increases until mechanical failure (rupture). We have adopted a van der Waals radius ($1.7$\AA) for the carbon atom in order to compute the von Mises stress.
\label{stressvonmises}}
\end{figure}

\begin{figure*}
\centering
\mbox{\subfigure[]{\includegraphics[width=0.47\linewidth]{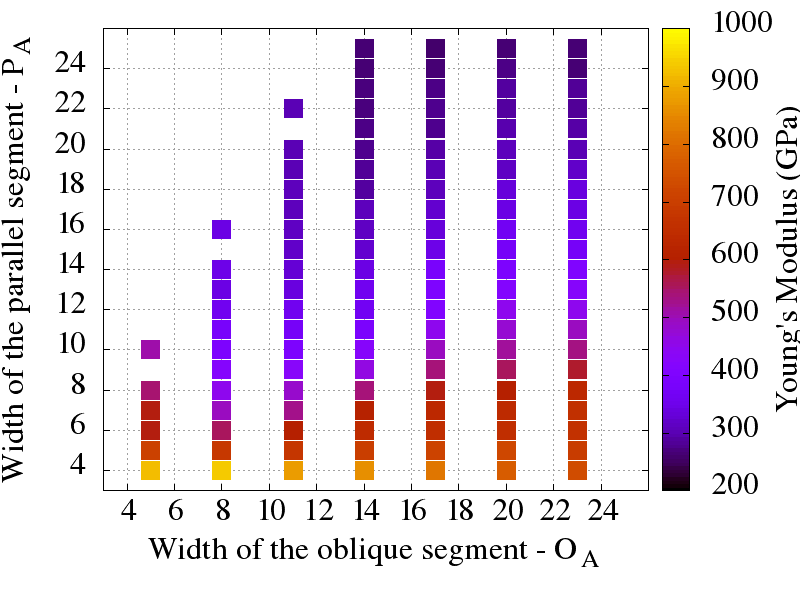}}\quad
\subfigure[]{\includegraphics[width=0.47\linewidth]{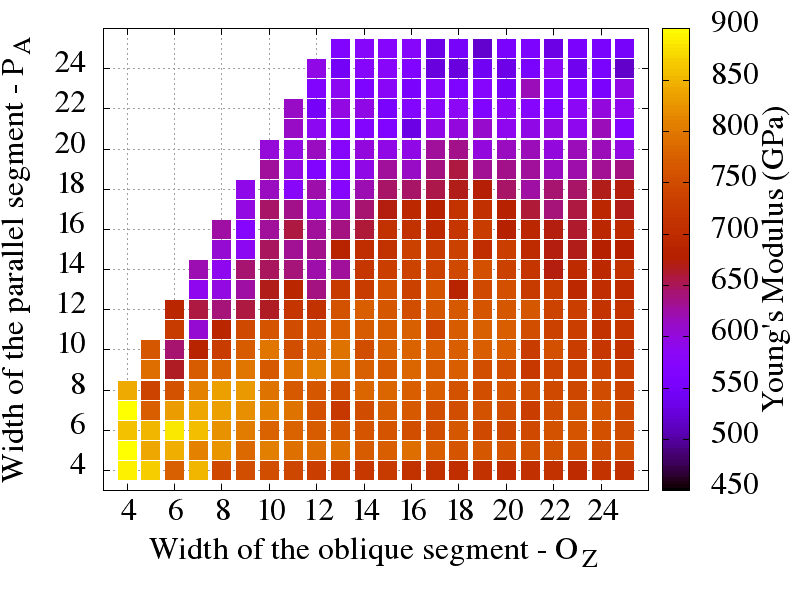} }}\\
\mbox{\subfigure[]{\includegraphics[width=0.47\linewidth]{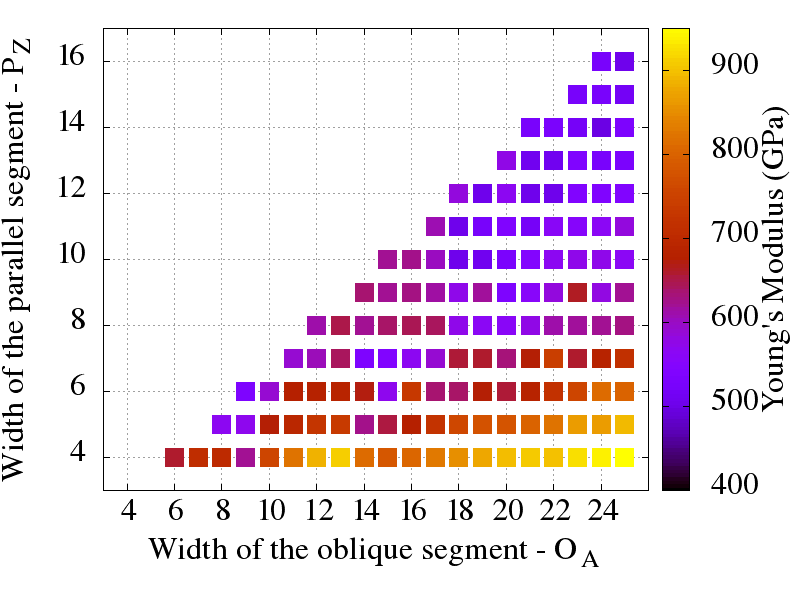}}\quad
\subfigure[]{\includegraphics[width=0.47\linewidth]{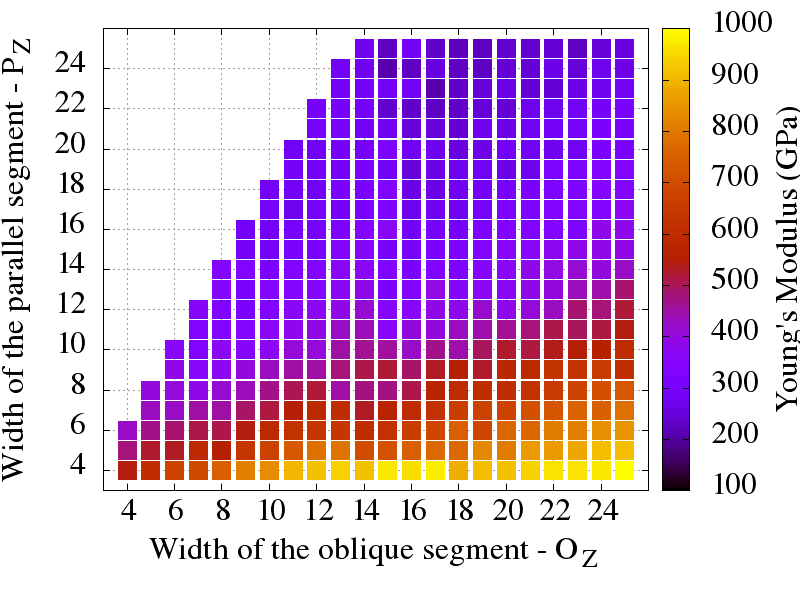} }}
\caption{\textit{Dependence of the Young's modulus on the width of the parallel and oblique segments ($P_\alpha$ and $O_\beta$, respectively) for (a) $AA$, (b) $AZ$, (c) $ZA$ and (d) $ZZ$ GNW. In general, Young's modulus values decrease with an increase of the parallel segment. 
%The von mises stress values was estimated considering the carbon atoms as a sphere of constant radius equal to carbon van der waals radius and then dividing the in plane von mises stress by the volume of all the carbon atoms.
}}
\label{youngall}
\end{figure*}

We then proceed with the analyses of the GNW mechanical properties under strain. As the strain increases, all levels of corrugations start to decrease, disappearing when the stress begins to increase in a linear regime. From this point, stress begins to increase linearly, characterizing the elastic behaviour of the material. From this linear regime we can calculate the Young's modulus values. In this regime, there is no plastic deformation, \textit{i.e.}, the structure returns to its initial configuration if the strain is removed. As strain continues to be increase, stress increases and the structure presents a non-linear behaviour until it reaches the ultimate strength point $\sigma_c$. At this point, the fracture process starts to occur and the stress values fast drops to zero. Representative stress-strain curves for one structure of each family are shown in Fig. \ref{stressall} (a), where each regime can be easily identified.

During the stretching process, we observe that the von Mises stress accumulates in a central line along the longitudinal direction of the GNW and reaches its maximum value at the inner corners of the structure, as can be seen in Fig. \ref{stressvonmises}. This can be explained by the fact that the regions far from this central line can easily relieve stress due to their unconstrained boundaries. Only the central line is constrained on both sides along the direction of the applied strain. The inner corners accumulate even more stress than the rest of the central line because of the force imbalances caused by absence of neighbors in one direction, which eliminates the internal reaction forces that otherwise would distribute the stress.

%**This can be explained by the fact that the stress propagation occurs along and in the direction defined by the structure bonds. As the number of bonds in the corners region is lower the stress will propagate in the direction of the remaining bonds in the center of the corner region and an increase in the local stress will be observed.** \emph{\textbf{. This behavior has already described in the book of XXX and YYY \cite{??}. ***Colocae a citação do livro depois pra eu dar uma lida lá, desse jeito aqui me parece meio confuso - tb achei confuso, o timoteo q botou isso ae - te mandei no email a explicacao}} 
%The distribution along the structure for each of the GWN families can be observed in the supplementary material figure ***X***.

It is a well-know fact that defects can locally weaken the material, favoring fracture to occur in that region and at lower stress values than the observed for the corresponding pristine material \cite{buehler2008atomistic,inglis1997stresses}. As previously above mentioned, the GNW shape can be considered as topological graphene intrinsic defects, so it should be expected that the GNW ultimate strength should be lower than that of graphene and that the fracture should occur at those regions. That is exactly what we observed in our simulations, since the strongest GNW is still weaker than pristine graphene, with cracks being usually formed at the vertices of the wiggles. This behavior was consistent for all distinct GNW families investigated here.

The nature of the fracture process strongly depends on the shape of the structures, ranging from brittle to super-ductile. A ($4_A,20_A$) GNW, for example, presents highly brittle behavior, with stress abruptly falling to zero after the fracture starts, as shown in Fig. \ref{stressall} (b). On the other hand, for a $(25_A,14_A)$ GNW, a ductile behaviour is observed and complete rupture is only obtained for strain values larger than $0.3$. In this case, a more complex process of stress alleviation can be observed, with several successive steps, as shown in Fig. \ref{stressall} (b), differing from the abrupt decrease observed for graphene \cite{ovid2013mechanical} and ($4_A,20_A$) GNW structure. These steps in the stress/strain curve are a consequence of an unravel-like process in the bond breaking that leads to a super-ductile behavior. In general, super-ductile behaviour was observed for structures with large parallel segments, \textit{i.e.}, large $P_\alpha$ values. This phenomenon was more pronounced for the $ZZ$ family, with some structures reaching final strain values as high as $0.5$. The final strain values reported here are larger than that ones previously reported by Xu \textit{et al.} \cite{Xu2013} for graphene with defects, especially when comparing with structures of the $ZZ$ family, as can be seen in Fig. 4 of the supplementary material.

%About the fracture process, we have observed different behaviors depending on the structures. For example, considering the ($4_A,20_A$) GNW when the fracture begins the stress decreases abruptly to zero (figure \ref{stressall} (b)). In the other hand, for $(25_A,14_A)$ GNW, a ductile behaviour is observed with a total fracture only for strain values higher than $0.3$. During the fracture process of the $(25_A,14_A)$ GNW, intermediate falls in the stress are observed (figure \ref{stressall} (b)) differing from the abrupt decrease observed for graphene \cite{} and ($4_A,20_A$) GNW structure. These falls in the stress comportment are related to the break of the bonds in a unravel-like process during the fracture dynamic. In general, super-ductility behaviour was observed for structures with high parallel segment, {\it i.e.} high $P_\alpha$ values, where some structures reach strain values up to $0.5$. The final strain values reported here are high than that ones previously reported for defected graphene by Xu et. al.\cite{Xu2013}, as can be seen in the supplemental figure 4. We can see a huge difference from the ZZ family to the others, reaching higher maximum strain due to its unravel-like fracture process.

We observed that GNW fractures usually propagate along the oblique directions. In this way, $AA$ and $ZA$ GNW families present armchair edges along the fracture, while $AZ$ GNW family present zigzag edges. On the other hand, an armchair fracture is more common for the $ZZ$ family.
Snapshots of the full fracture process of a ($25_A,14_A$) GNW can be seen in Fig. \ref{stressvonmises}. The corresponding videos depicting the whole dynamics for ($25_A,14_A$) and ($4_A,20_A$) GNW can be found in the supplementary material. 

Considering all GNW families, the Young's modulus and the ultimate strength $\sigma_c$ values decrease with the increase of the $P_\alpha$ parameter. However, we could not verify a consistent dependence on the $O_\beta$ parameter. The values of the Young's modulus for all investigated GNW structures are presented in Fig. \ref{youngall}. The ultimate strength, $\sigma_c$, follows a similar trend, which can be seen in Figs. 3 and 5 of the supplementary material. Temperature effects are negligible on the Young's modulus values (see Fig. 6 of the supplementary material), as they remain virtually unchanged from $10K$ to $300K$. On the other hand, ultimate strength values are very sensitive to temperature changes, sharply decreasing with increasing temperature.

Among the four families, the Young's modulus for the $AA$ and $ZZ$ families can reach values from $100$ to $1000$ GPa (Fig. \ref{youngall} (a) and (d)), while for the $AZ$ and $ZA$ families the values range from $400$ to $900$ GPa (Fig. \ref{youngall} (b) and (c)). The ultimate strength values range from $20$ to $100$ GPa for $AA$, $40$ to $110$ for $AZ$, $40$ to $90$ for $ZA$, and  $20$ to $100$ for $ZZ$. The very large range of Young's modulus, final strain and ultimate strength values provides great tunability to the GNW mechanical properties, enabling them to be tailored for specific applications.

\begin{figure}
\centerline{\includegraphics[width=1.0\linewidth]{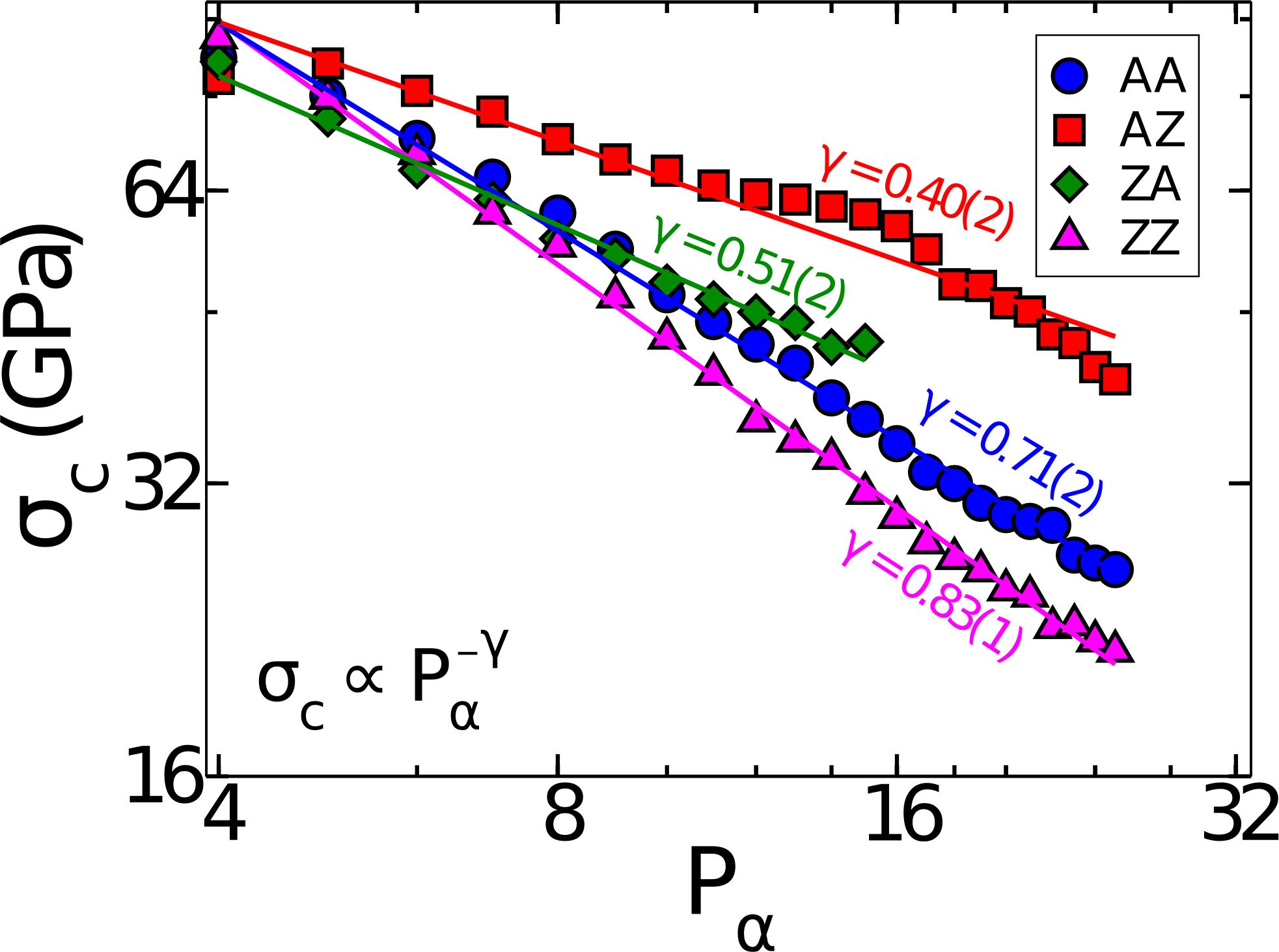}}
\caption{Ultimate strength (US) as a function of $P_\alpha$ keeping $O_\beta$ parameter constant at $O_\beta=23$ for all the GNW families. It can be seen that the ultimate strength follows a power law, decreasing while the $P_\alpha$ parameter grows.
\label{stresspotencia}}
\end{figure}

Another interesting result is that the ultimate strength values depend on the $P_\alpha$ parameter in the form of a power law, as can be seen in Fig. \ref{stresspotencia}. For this analysis, the $O_\beta$ parameter was kept constant at $O_\beta=23$. The exponents of the power law depend on the GNW family. The $ZZ$ and $AA$ families present the largest exponents: $\gamma=0.83(1)$ for $ZZ$ family and $\gamma=0.71(2)$ for $AA$ family, followed by the $ZA$ family with $\gamma=0.51(2)$ and $AZ$ family with $\gamma=0.40(2)$. The power law regressions may be very useful to estimate the ultimate strength of uncalculated GNW structures.

The fact that $AZ$ and $ZA$ families have larger ultimate strength may be attributed to their larger opening angle ($120^{\circ}$) when compared to $AA$ and $ZZ$ families ($60^{\circ}$). As previously attested by Carpinteri \cite{Carpinteri20051254}, structures with re-entrant corners get stronger when the mass of the structure decreases, \textit{i.e.} the angle of the corner increases.

\section{Conclusions}
\label{conclusions}
We have investigated through fully atomistic reactive molecular dynamics simulations the mechanical properties of graphene (graphitic) nanowiggles under strain for different temperatures. We calculated the Young's modulus, ultimate strength and stress distribution during the stretching process, as well as the fracture patterns for different GNW families. 

GNW are shown to present very diverse mechanical properties, which strongly depend on their shape. In special, a super-ductile behavior was observed for structures with large values of $P_\alpha$, while brittle behavior was the general trend for the other GNW. The fracture dynamics for the super-ductile structures present an unravel-like process of the hexagonal rings near to the inner corners. For some structures, complete fracture happens for strain values as high as 0.5. 

Young's modulus and ultimate strength values range from $100$ to $1000$ GPa and $20$ to $110$ GPa, respectively. Also, a power law dependence on the width of the parallel segment, $P_\alpha$, was found for the ultimate strength. The wide range of values for Young's moduli, final strains and ultimate strengths and the distinct fracture behaviors provide GNW with an unusual and highly promising level of design versatility. The direct dependence of these properties on the shape of the GNW creates an easily accessible path for tuning. Combining this rich mechanical behavior with their previously reported tunable electric and magnetic properties \cite{giraoprl}, makes GNW one of the most exciting and attractive novel structures to be exploited as the basis for nanodevices and advanced functional materials.

%%%%%%%%%%%%%%%%%%%%%%%%%%%%%%%%%%%%%%%%%%%%%%%%%%%%%%%%%%%%%%%%%%%%%%%%%%%%%
\section*{Acknowledgments}
This work was supported in part by the Brazilian Agencies CNPq, CAPES and FAPESP. The authors would like to thank the Center for Computational Engineering and Sciences at Unicamp for financial support through the FAPESP/CEPID Grant $2013/08293-7$. N.M.P. is supported by the European Research Council PoC 2015 ``Silkene" No. 693670, by the European Commission H2020 under the Graphene Flagship Core 1 No. 696656 (WP14 ``Polymer Nanocomposites") and under the Fet Proactive ``Neurofibres" No. 732344.
%%%%%%%%%%%%%%%%%%%%%%%%%%%%%%%%%%%%%%%%%%%%%%%%%%%%%%%%%%%%%%%%%%%%%%%%%%%%%
%%%END OF MAIN TEXT%%%

%\section*{References}

\bibliographystyle{plain}
\bibliography{bib}

\end{document}